\documentclass[letterpaper, 10 pt, conference]{ieeeconf}  

\IEEEoverridecommandlockouts                              
\usepackage{amsmath} 
\usepackage{array}
\usepackage{balance}
\usepackage{rotating}

\usepackage{verbatim}
\usepackage{graphicx}
\usepackage{eurosym}
\usepackage{}
\title{\LARGE \bf
Integrated Process Planning and Scheduling\\ in Commercial Smart Kitchens}

\author{Piotr Dziurzanski, Shuai Zhao and Leandro Soares Indrusiak\\
University of York, Deramore Lane,
York, YO10 5GH, UK.\\
{\tt \{firstname.lastname\}@york.ac.uk}\\
}

\begin{document}

\maketitle
\thispagestyle{empty}
\pagestyle{empty}

\begin{abstract}
This paper describes the possibility of applying a generic, cloud-based Optimisation as a Service facility to food cooking planning and scheduling in a commercial kitchen. We propose a  chromosome encoding and customisation of the classic MOEA/D multi-objective genetic algorithm. The applicability of the proposed approach is evaluated experimentally for two scenarios different with respect to the number of cooking appliances and the amount of the ordered food. The proposed system managed to determine the trade-offs between cooking time, energy dissipation and food quality.

\end{abstract}

\section{Introduction}

\begin{figure}[b]
\centering
\includegraphics[width=.8\columnwidth]{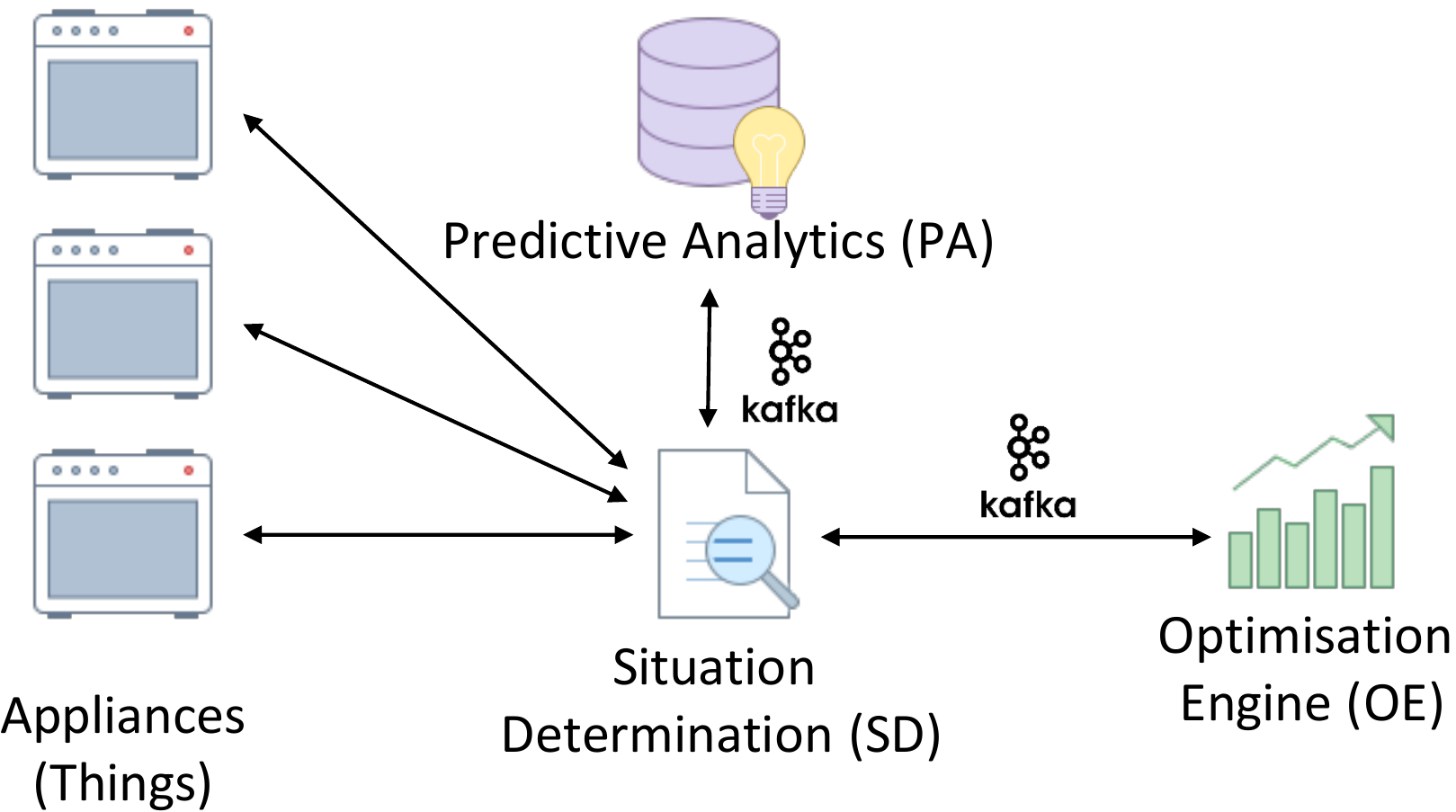}
\caption{Overall architecture of the proposed system}\label{fig:overallArchitecture}
\end{figure}

The term fourth industrial revolution, often abbreviated to Industry 4.0, is used to refer to the current trend of automation in manufacturing technologies. One of the promises of Industry 4.0 is related to facilitating of production of small batches of highly customised commodities \cite{Wang2016}. Even if not 100\% correct from the orthodox standpoint, preparing dishes in a restaurant may be viewed to be consistent with the primary assumptions of Industry 4.0. The orders are released at random moments, the batches are extremely short (up to a number of people in a group) and highly customised (e.g., blue, rare, medium rare, medium or well-done steaks). Consequently, a chef is often reported as one of the most stressful jobs \cite{Gibbons2007}. 
Digital gastronomy aims to mitigate the chef life by enhancing traditional cooking with new capabilities rather than replacing the chef with an autonomous machine
\cite{Mizrahi2016}. Following this trend, modern technologies are increasingly more popular in restaurants. At the moment, the Internet of Things (IoT) is used to monitor the equipment that cooks, cleans or stores food \cite{Kiesel18}. However, in the near future technology will be probably applied to more sophisticated tasks. For example, the recent progress in developing so-called electronic tongues and noses can facilitate the automation of the process of food samples' quality estimation \cite{Podrazka18}. Such sensors can be used to fill the gap between recipes and actual cooking activities, identified in \cite{Sato2014}. Even Kinect-style cameras can be applied for fine-grained recognition of kitchen activities \cite{Lei2012}. Not only the sensors but also the actuators in smart kitchen appliances can act as things connected to the Internet, as for example a recently presented smart oven from Electrolux in which the time duration or temperature can be remotely altered \cite{Electrolux18}. Even if a certain task in a cooking process has to be done manually (e.g., adding ingredients to a pot), it can be guided by a robot speech following some recipes \cite{Suzuki2012} or even a cognitive conversational agent connected to a smart fridge \cite{Angara2017}. However, those systems process one recipe at a time. This is in contrast to \cite{Hamada2005}, where the cooking process has been treated as an optimisation problem. That system applied the list scheduling algorithm to minimise the food time preparation and maximise food quality, benefiting from the fact that some actions can be executed in parallel, reducing the cooking duration.

The above observations encourage us to apply a generic factory reconfiguration system, described in \cite{Dziurzanski2019EvoApp}, to food preparation in restaurants. In particular, similarly to a smart factory, a kitchen can receive a new order at any time. For such order, a process planning and scheduling need to be performed with no delay \cite{Dziurzanski2019ICIT}. Process planning and scheduling are required to be re-executed in case any unexpected event occurs in the kitchen, for example, a failure of a device (treated as a smart thing) is detected \cite{Wan19}. If a number of devices in a kitchen is considerable, process planning and scheduling will have similar computation cost as in a smart factory, which is rather significant \cite{Sobeyko2017}. Between subsequent executions of the process planning and scheduling, the computational power is not needed. Consequently, the workload related to the process planning and scheduling in a restaurant follows the on-and-off workload pattern. When using the process planning and scheduling approach that can be processed in a distributed way, for example the island model of a Genetic Algorithm (GA) similarly to \cite{Zhao19}, the workload satisfies all criteria for suitability for public cloud provided in \cite{Geyer2012}, namely an unpredictable load, different computational power requirements at different time intervals and horizontal scalability.

The rest of this paper is organised as follows. Section \ref{sec:generic} outlines our generic service for optimisation of smart factories. Application of this service to a smart kitchen is described in Section \ref{sec:applying}. In Section \ref{sec:experimentalResults}, experimental evaluation of the proposed approaches is carried out. Section \ref{sec:ConcludingRemark} concludes the paper.

\section{Generic service for smart factories re-configuration}\label{sec:generic}

The class of optimisation problems analysed in this paper concerns integrated process planning and scheduling in smart kitchens performed on demand. The optimisation is carried out by the module named Optimisation Engine (OE), which is a part of the larger system presented in Figure \ref{fig:overallArchitecture}. The operation of this system is triggered by the data ingested from assorted devices (things) such as smart hobs. This data is collected by the Situation Determination (SD) module, which derives the current situation of resources, products and processes, using a custom use-case situation model based on a common situation model. SD monitors raw data provided by things and the outcomes of the Predictive Analytics (PA) module and then determines the kitchen current state. In case any relevant change of the kitchen state is detected, the process plan and schedule is recomputed.

\begin{figure}
\centering
\includegraphics[width=\columnwidth]{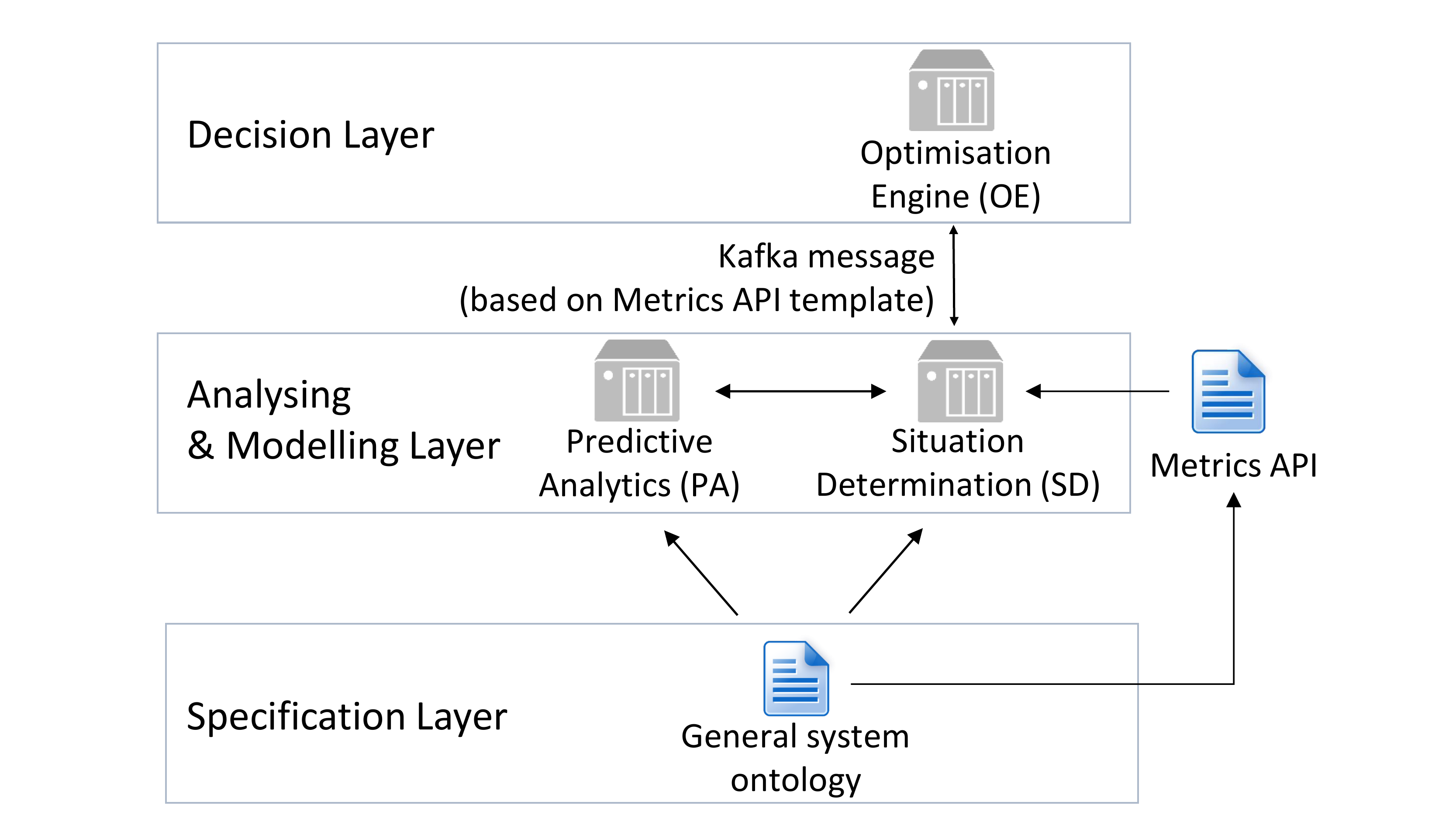}

\caption{Layers in the proposed approach}\label{fig:designFlow}
\end{figure}

In the proposed system, the popular streaming platform named Kafka has been applied to the communication between modules. The messages sent via Kafka follow a well-defined textual protocol named Metrics API, which defines three types of numeric or nominal values: key objectives to be optimised, control metrics that can be mutated to obtain various candidate solutions and observable metrics informing about situations relevant to the optimisation process, such as unavailability of a certain resource. This system architecture is compliant with the scheme proposed in \cite{Alsafi2010}, where three-layers of an intelligent factory reconfiguration system have been identified. The lowest layer, named \emph{Specification layer}, includes the knowledge description regarding the factory based on General system ontology. The middle layer, \emph{Analysing \& modelling layer}, includes both SD and PA modules. OE operates in the topmost layer which is named \emph{Decision layer}. These layers are visualised in Figure \ref{fig:designFlow}. From this layered architecture, it follows that the appropriate operation of both the SD and PA modules are crucial for performing effective optimisation. However, the detailed description of these two modules is out of the scope of this paper.

The overall specification of reconfiguration capability is most readily explained via division into two component parts: the Metrics API and Optimisation Engine. The Metrics API provides a complete configuration description for the variables associated with the food order and kitchen temporal state. The elements of the configuration data schema are termed metrics, i.e. either measurable physical values corresponding to appliance sensors or else key objective (quality) measures derived from these. The chief functionality of OE is to generate a food cooking plan and schedule in response to reconfiguration requests issued by SD. The quality of the candidate plans and schedules is determined by the objective function. This function is generated automatically based on a kitchen configuration and applies a digital twin of the corresponding smart kitchen specified with Interval Algebra \cite{Dziurzanski2019ICIT}.

The optimisation used by OE is based on \cite{Dziurzanski2019EvoApp}. There are given a set of recipes, a set of resources and an order. OE assigns resources and priorities to a multisubset (i.e. a combination with repetitions) of recipes so that the total processing time (makespan) and energy are minimised and the food quality is the highest. In genetic algorithms, candidate solutions are treated as individuals. During the optimisation process, these individuals are evolved using a set of bio-inspired operators such as selection, crossing-over and mutation. The solution to the problem considered in this paper can be then described with a chromosome whose odd genes identify the recipe to be applied and the even genes denote the priority of that recipe instance. The aim of introducing priorities is to determine the processing orders of recipes allocated to the same resource and thus to determine the temporal scheduling. This ordering does not change the amount of cooked food but can influence the makespan.

The problem analysed in this paper is characterised with multi-objective criteria. End-users should be then informed about a wide set of Pareto-optimal solutions to select the final solution based on their knowledge of the problem. The set of the alternative solutions presented to the end-users should be then diverse and, favourably, distributed over the entire Pareto front. This expectation is in line with the properties of the MOEA/D algorithm proposed by Zhang and Li in \cite{Zhang2007}, which is then used in OE.

\section{Applying generic re-configuration service for smart kitchen}\label{sec:applying}

\begin{figure}
\includegraphics[width=\columnwidth]{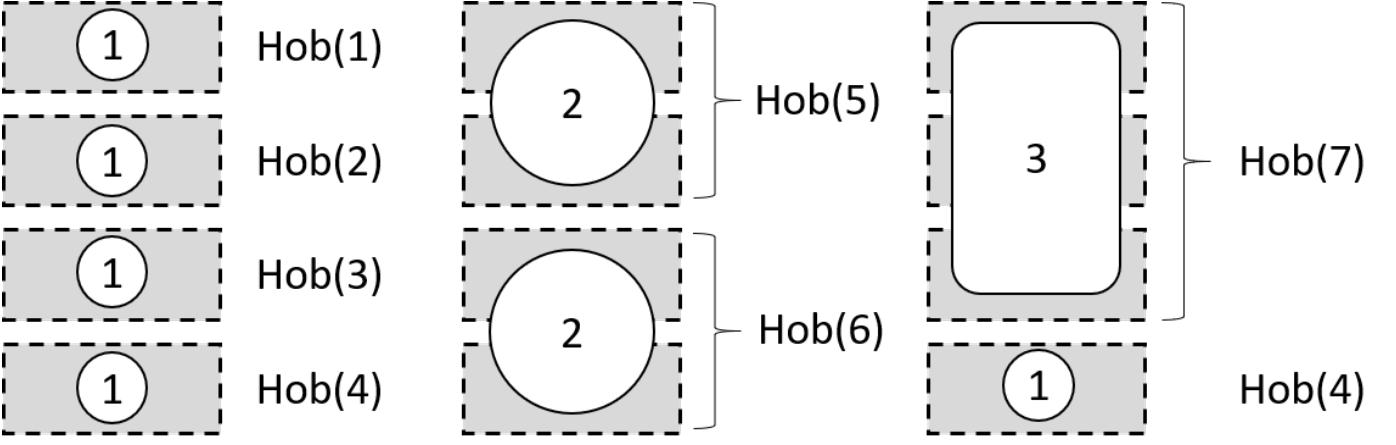}

\caption{Example of possible configuration of four cooking zones to heat pots of different sizes}\label{fig:23}

\end{figure}

The optimisation process is executed after sending a serialised configuration to a predefined Kafka topic. The configuration includes the list of dishes to be cooked together with the list of the recipes and the parameters of the available hobs. An example list of recipes used in this paper is provided in Table \ref{tab:Appendix}. For example, the list of compatible cooking zones (e.g. of the area sufficient for a certain pot or food amount) is provided in the domain of a controlled metric type, which is shown in Figure \ref{fig:22} for two examples related to boiled water. The resource name is composed of two parts: the actual cooking zone name and the pot type. Each recipe can be executed a number of times, thus the recipe name is followed by an instance index (e.g. suffix 0 in Boiled water A 0). As it is shown in this figure, the first recipe, Boiled water A 0 can be executed using Pot(1) and cooking zones 1-4 on hob named Hob, or not selected to be executed (i.e., No allocation). In the assumed hob, a few cooking zones can be used simultaneously to cook a dish in a larger pot, as shown in Figure \ref{fig:23}. In this figure, four cooking zones can be used with pot(1) (i.e., the circles on the hob) independently (left column), or two upper or two lower cooking zones can be combined and used with pot(2) (middle column). Notice, that using two middle cooking zones in such a way is impossible. Finally, three upper zones can be used simultaneously with pot(3) (right column). In the configuration, all cooking zones Hob(1)-Hob(7) are provided as separate resources, but the fact that certain resources cannot be used simultaneously (e.g. Hob(1) and Hob(5)) is defined using mutual exclusiveness of resources as explained in \cite{Dziurzanski2019ICIT}.

\begin{figure}
\footnotesize
\begin{verbatim}
ControlledMetricType[name=Boiled water A 0,
allocation,valueType=ValueType.Nominal[name=
Boiled water A 0 allocation type ,values={Hob(1)
Pot(1), Hob(2) Pot(1), Hob(3) Pot(1), Hob(4) Pot(1),
No allocation},typ=NOMINAL]

ControlledMetricType[name= Boiled water B 0
allocation,valueType=ValueType.Nominal[name=
Boiled water B 0 allocation type ,values={Hob(5)
Pot(2), Hob(6) Pot(2), No Allocation},typ=NOMINAL],
units=n/a]
\end{verbatim}
\caption{Two example controlled metrics}\label{fig:22}
\end{figure}

The observable metrics are used to denote the availability of certain resources. Due to the compound naming structure of these resources, both a certain cooking zone or a certain pot type can be signalled as being unavailable. For example, temporal lack of pot(1) should result in the unavailability of all resources that are used in combination with that pot. The first example in Figure \ref{fig:24} signals unavailability of a certain cooking zone. The second and third observables inform OE about the cooking time of a certain recipe (here: Boiled water A) with a certain cooking zone (here: Hob(1)) and pot (here: Pot(1)), as determined by the predictive analytics or situation determination modules. This cooking time is taken into consideration when a schedule is determined. Finally, the recipe quality can be updated based on some user feedback. 

\begin{figure}
\footnotesize
\begin{verbatim}
ObservableMetricType[name= Hob(6) availability,
valueType=ValueType.Integer[min=0,max=0,typ=INT],
units=n/a,sampleRate=SampleRate.EventDriven[]],

ObservableMetricType[name=Boiled water A Hob(1)
Pot(1) start, valueType=ValueType.Integer[min=0,
max=0,typ=INT]

ObservableMetricType[name=Boiled water A Hob(1)
Pot(1) end, valueType=ValueType.Integer[min=40,
max=40,typ=INT]
\end{verbatim}
\caption{Two example observable metrics in the Electrolux use case}\label{fig:24}
\end{figure}

\begin{figure}
\footnotesize
\begin{verbatim}
Optimisation took: 19.942 seconds
Schedule: Status: Succeeded.

Hob(2) Pot(1) -> [
Rice A 1_1 [35,37),
Rice A 1 [103,128),
Beef A 2_1 [188,200),
Beef A 2 [692,812),
DependentSetUp from Beef A
    to Boiled Water A [812,822),
Boiled Water A 0 [1094,1109),
DependentSetUp from Boiled Water A
    to Pasta A [1109, 1119),
Pasta A 0_1 [1297,1299),
Pasta A 0 [1299,1319)
]
…
makespan: 28:57.00
\end{verbatim}
\caption{Extract from an example OE report for the Electrolux use case}\label{fig:25}
\end{figure}

\section{Experimental results}\label{sec:experimentalResults}

In the considered scenario, the following amounts of food are required to be cooked: Boiled water - 5000g, Pasta - 1000g, Rice - 1500g, Meat (beef) - 1000g, Vegetable (potatoes) - 1000g, Mushrooms (with oil) - 500g.
These values are provided to OE via Kafka and then the optimisation process is executed. An example report of the optimisation process is presented in Figure \ref{fig:25}. As shown in the figure, the tasks are executed concurrently on all available resources, and the finish time of the last task indicates the makespan of this optimisation procedure. Note the start and end time of each task is relative to the starting time point 0. For the production that requires certain pre-cooking, subtasks are introduced and are executed before cooking the required production (i.e., task Rice A 1\_1 for cooking rice using task Rice A 1). In addition, dependent setup is required when different production is executed on the same resource (see task DependentSetUp).





\begin{figure}
\includegraphics[width=\columnwidth]{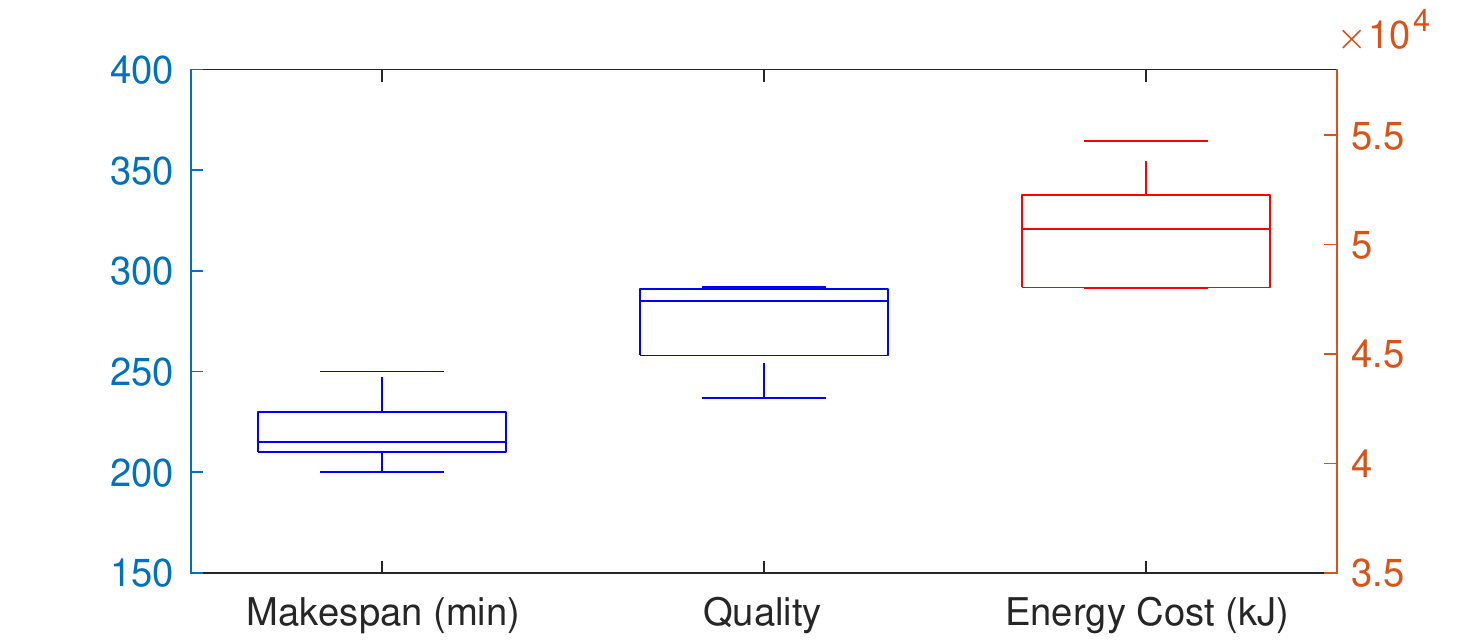}

\caption{Pareto front approximation (Makespan, Quality and Energy Cost) while cooking recipes from Table \ref{tab:Appendix}}\label{fig:oneHub}

\end{figure}




\begin{figure}
\includegraphics[width=\columnwidth]{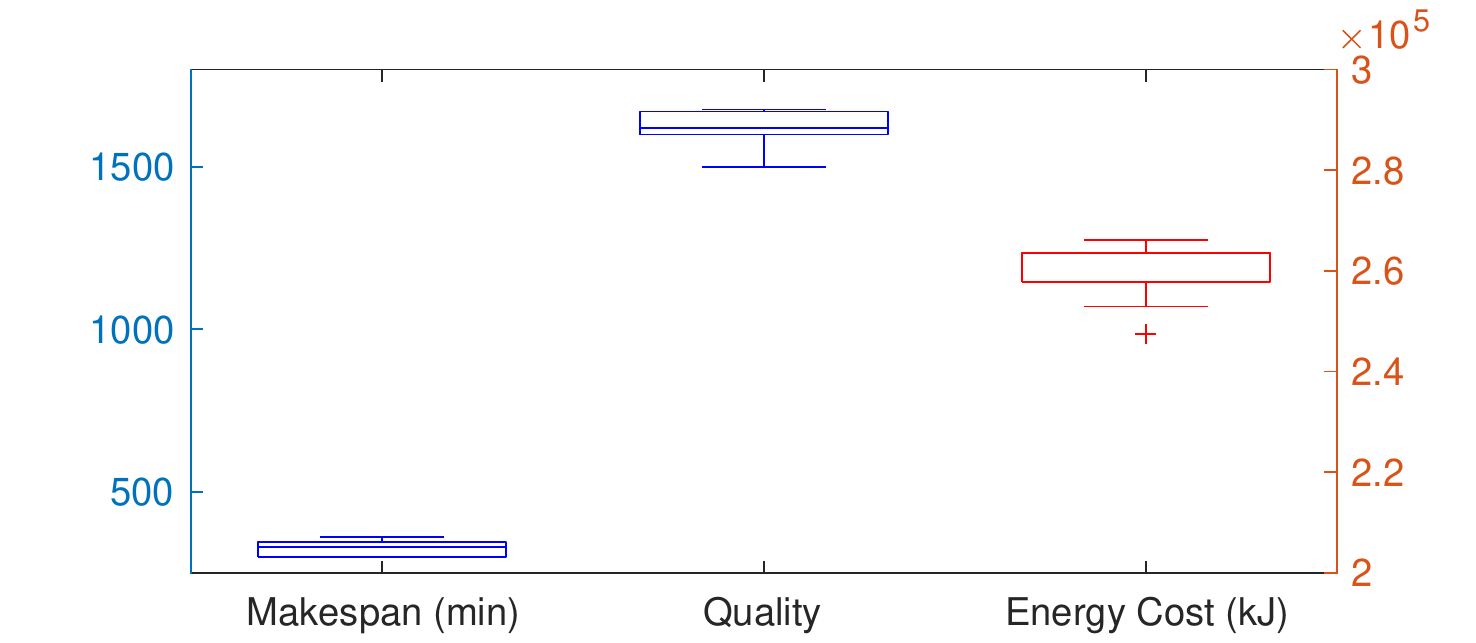}

\caption{Pareto front approximation (Makespan, Quality and Energy Cost) while cooking scaled recipes from Table \ref{tab:Appendix} in a commercial scale scenario}\label{fig:fourHub}

\end{figure}

The trade-off between three conflicting objectives, makespan, energy cost and deficiency (the reverse of quality), has been investigated for the example scenario and visualised in Figure \ref{fig:oneHub} in a form of a Pareto front approximation. The chart demonstrates that the decreasing of makespan is obtained via sacrificing the quality (i.e., increasing deficiency). Note, during the experiment, we noticed that it is not necessary that a higher makespan must lead to a lower deficiency value (which indicates a higher food quality). This is because the recipes can be executed in parallel (based on the allocation decision during optimisation). Given the same recipes, executing in parallel can decrease the makespan but the deficiency objective will remain unchanged. Therefore, during this evaluation, we often observed new optimisation results that contain makespan and deficiency values that are both lower than certain previous optimisation results. Such a phenomenon also makes the final number of optimisation results low, as optimisation results that are strictly dominated will be removed from the Pareto front approximation. However, in general, we demonstrate the trend where the deficiency metric value is decreased (i.e., better quality is obtained) while the makespan is increasing.
In addition,
The trade-off between makespan and energy cost, has been
investigated for the example scenario and demonstrated in Figure \ref{fig:oneHub} in the form of a Pareto front approximation. The chart indicates that the decreasing of makespan can be achieved via consuming more energy. 

The proposed method is applicable rather to large restaurants or company cafeterias where trade-offs between conflicting objectives such as energy, makespan and deficiency, needs to be investigated rather than to single hobs treated as home appliances. With a larger scenario, both in terms of the number of available resources and the ordered food, it needs to be considered to demonstrate the scalability of the proposed approach. Hence in the second scenario, the presence of 4 hobs from Figure \ref{fig:23} is assumed. In addition, the amount of the ordered food is 4 times larger than in the previous case. 

\begin{sidewaystable*} 
\footnotesize
\centering
\caption{Parameters of food cooking using various cooking zones, pots and recipes}\label{tab:Appendix}
\begin{tabular}{l|lllllllll}
Food type                    & Recipe name & Predecessor    & Amount (g) & Cooking zone                   & Pot type & Energy (kJ) & Cooking time (min) & Monetary cost (€) & Deficiency \\ \hline
Boiled water                 & A           & -              & 1000       & Hob(1), Hob(2), Hob(3), Hob(4) & Pot 1    & 350         & 15                 & 0.03              & 5           \\
                             & B           & -              & 2000       & Hob(5), Hob(6                  & Pot 2    & 1400        & 10                 & 0.12              & 8           \\
                             & C           & -              & 3000       & Hob(7)                         & Pot 3    & 3150        & 5                  & 0.27              & 11          \\ \hline
Pasta                        & A           & Boiled water A & 100        & Hob(1), Hob(2), Hob(3), Hob(4) & Pot 1    & 840         & 30                 & 0.021             & 2           \\
                             & B           & Boiled water A & 100        & Hob(1), Hob(2), Hob(3), Hob(4) & Pot 1    & 770         & 25                 & 0.018             & 9           \\
                             & C           & Boiled water B & 200        & Hob(5), Hob(6)                 & Pot 2    & 1120        & 20                 & 0.021             & 14          \\
                             & D           & Boiled water B & 200        & Hob(5), Hob(6)                 & Pot 2    & 1190        & 15                 & 0.018             & 19          \\
                             & E           & Boiled water C & 300        & Hob(7)                         & Pot 3    & 1520        & 10                 & 0.021             & 22          \\
                             & F           & Boiled water C & 300        & Hob(7)                         & Pot 3    & 1590        & 5                  & 0.018             & 25          \\ \hline
Rice                         & A           & Boiled water A & 200        & Hob(1), Hob(2), Hob(3), Hob(4) & Pot 1    & 1260        & 50                 & 0.045             & 7           \\
                             & B           & Boiled water A & 200        & Hob(1), Hob(2), Hob(3), Hob(4) & Pot 1    & 1400        & 45                 & 0.039             & 15          \\
                             & C           & Boiled water B & 400        & Hob(5), Hob(6)                 & Pot 2    & 1610        & 40                 & 0.045             & 19          \\
                             & D           & Boiled water B & 400        & Hob(5), Hob(6)                 & Pot 2    & 1750        & 35                 & 0.039             & 22          \\
                             & E           & Boiled water C & 600        & Hob(7)                         & Pot 3    & 1960        & 15                 & 0.045             & 28          \\
                             & F           & Boiled water C & 600        & Hob(7)                         & Pot 3    & 2100        & 13                 & 0.039             & 33          \\ \hline
Meat (beef)                  & A           & Boiled water A & 250        & Hob(1), Hob(2), Hob(3), Hob(4) & Pot 1    & 4550        & 120                & 0.27              & 5           \\
                             & B           & Boiled water A & 250        & Hob(1), Hob(2), Hob(3), Hob(4) & Pot 1    & 6650        & 110                & 0.18              & 9           \\
                             & C           & Boiled water B & 500        & Hob(5), Hob(6)                 & Pot 2    & 6900        & 90                 & 0.27              & 12          \\
                             & D           & Boiled water B & 500        & Hob(5), Hob(6)                 & Pot 2    & 7000        & 85                 & 0.18              & 16          \\
                             & E           & Boiled water C & 750        & Hob(7)                         & Pot 3    & 7350        & 60                 & 0.27              & 21          \\
                             & F           & Boiled water C & 750        & Hob(7)                         & Pot 3    & 7550        & 55                 & 0.18              & 27          \\ \hline
Vegetable           & A           & Boiled water A & 200        & Hob(1), Hob(2), Hob(3), Hob(4) & Pot 1    & 1750        & 42                 & 0.066             & 3           \\
(potatos)                             & B           & Boiled water A & 200        & Hob(1), Hob(2), Hob(3), Hob(4) & Pot 1    & 1890        & 40                 & 0.06              & 11          \\
                             & C           & Boiled water B & 400        & Hob(5), Hob(6)                 & Pot 2    & 2100        & 32                 & 0.066             & 19          \\
                             & D           & Boiled water B & 400        & Hob(5), Hob(6)                 & Pot 2    & 2240        & 30                 & 0.06              & 23          \\
                             & E           & Boiled water C & 600        & Hob(7)                         & Pot 3    & 2450        & 22                 & 0.066             & 26          \\
                             & F           & Boiled water C & 600        & Hob(7)                         & Pot 3    & 2590        & 20                 & 0.06              & 31          \\ \hline
Cooking  & A           & -              & 200        & Hob(1), Hob(2), Hob(3), Hob(4) & Pot 1    & 700         & 38                 & 0.072             & 11          \\
with oil                              & B           & -              & 200        & Hob(1), Hob(2), Hob(3), Hob(4) & Pot 1    & 840         & 36                 & 0.06              & 16          \\
(mushroom)                             & C           & -              & 300        & Hob(5), Hob(6)                 & Pot 2    & 910         & 25                 & 0.09              & 19          \\
                             & D           & -              & 300        & Hob(5), Hob(6)                 & Pot 2    & 1050        & 23                 & 0.078             & 20          \\
                             & E           & -              & 400        & Hob(7)                         & Pot 3    & 1120        & 12                 & 0.108             & 26          \\
                             & F           & -              & 400        & Hob(7)                         & Pot 3    & 1260        & 10                 & 0.096             & 29
\end{tabular}
\end{sidewaystable*}

\section{Concluding remark}\label{sec:ConcludingRemark}

In this paper, a real-world food cooking planning and scheduling problem in a commercial smart kitchen has been described. Its goal is not only to minimise the cooking time but also to minimise the energy dissipation and maximising the food quality via selecting recipes multisubset to be executed. A typical multi-objective genetic algorithms named MOEA/D has been used. The experiments have demonstrated the applicability of the proposed approach which has been able to determine the trade-offs between the conflicting objectives. The proposed algorithm is scalable enough to be applied to a relatively large kitchen and high quantity of production.

\section*{Acknowledgement}
The authors acknowledge the support of the EU H2020 SAFIRE project
(Ref. 723634).

The authors would like to thank Andrea De Angelis and Claudio Cenedese from ELECTROLUX AS for their support in defining the scenario described in this paper.

The icons used in this paper have been created by Icons8, https://icons8.com.

\balance


\bibliographystyle{IEEEtran}
\bibliography{smartKitchen}


\end{document}